\documentclass[letterpaper]{article} 
\usepackage{aaai21}  
\usepackage{amsmath,amsfonts,amssymb}
\usepackage{bm}
\usepackage{times}  
\usepackage{helvet} 
\usepackage{courier}  
\usepackage[hyphens]{url}  
\usepackage{graphicx} 
\usepackage{natbib}  
\usepackage{caption} 
\usepackage{cite}
\usepackage{textcomp}
\usepackage[shortlabels]{enumitem}
\usepackage{balance}
\usepackage{comment}
\usepackage{subcaption}
\usepackage{url}
\usepackage{color}

\usepackage{algorithm}
\usepackage{algpseudocode}
\usepackage{listings}
\usepackage{multirow}
\usepackage{titlesec}
\usepackage{etoolbox}
\usepackage{blindtext}
\usepackage{tcolorbox}

\newcommand{\inlinedComment}[2]
{\textcolor{#1}{\small\textbf{#2}}}
\newcommand{\lx}[1]{}
\newcommand{\n}[1]{}
\newcommand{\yu}[1]{}

\title{TreeCaps: Tree-Based Capsule Networks for Source Code Processing}

\author{
	Nghi D. Q. Bui \textsuperscript{\rm 1}
	Yijun Yu \textsuperscript{\rm 2}, 
	Lingxiao Jiang \textsuperscript{\rm 1}\\
}

\affiliations{
	\textsuperscript{\rm 1} School of Information Systems, Singapore Management University \texttt{\{dqnbui,lxjiang\}@smu.edu.sg} \\
	
	\textsuperscript{\rm 2} School of Computing \& Communications, The Open University, UK, \texttt{y.yu@open.ac.uk} \\

}

\begin{document}

	
	\maketitle
	
	\begin{abstract}


Recently program learning techniques have been proposed to process source code based on syntactical structures (e.g., Abstract Syntax Trees) and/or semantic information (e.g., Dependency Graphs). 
While graphs may be better at capturing various viewpoints of code semantics than trees, constructing graph inputs from code need static code semantic analysis that may not be accurate and introduces noise during learning. On the other hand, syntax trees are precisely defined according to the language grammar and easier to construct and process than graphs.
We propose a new tree-based learning technique, named TreeCaps, by fusing capsule networks with tree-based convolutional neural networks, to achieve learning accuracy higher than existing graph-based techniques while it is based only on trees.
TreeCaps introduces novel \textit{variable-to-static routing} algorithms into the capsule networks to compensate for the loss of previous routing algorithms. 
Aside from accuracy, we also find that TreeCaps is the most robust to withstand those semantic-preserving program transformations that change code syntax without modifying the semantics.
Evaluated on a large number of Java and C/C++ programs, TreeCaps models outperform prior deep learning models of program source code, in terms of both accuracy and robustness for program comprehension tasks such as code functionality classification and function name prediction. The implementation of TreeCaps is publicly available at \url{https://github.com/bdqnghi/treecaps}.

	\end{abstract}
	
\vspace{-8pt}
\section{Introduction} \label{sec:introduction}
Software developers often spend the majority of their time in navigating existing program code bases to understand the functionality of existing source code before implementing new features or fixing bugs~\citep{Xia2018,EDC2019,Britton2012}.
Learning a model of programs has been found useful for their tasks such as classifying the functionality of programs~\citep{Nix2017,dahl2013large,pascanu2015malware,rastogi2013catch}, predicting bugs~\citep{Yang2015,li2017software,li2018vuldeepecker,zhou2019devign}, translating programs~\citep{chen2018tree, Gu2017, BuiJY18, BuiYJ19, Bui2019}, etc.

It is common that adding semantic descriptions (e.g., via code comments, visualizing code control flow graphs, etc.) may enhance human understanding of programs and ease machine learning.
With the help of static code dependency analysis techniques~\citep{Nielson1999}, for example, Gated Graph Neural Networks (GGNN)~\citep{Li2016,fernandes2018structured, Allamanis2018} learn code semantics via graphs where edges are added between the code syntax tree nodes to indicate various kinds of dependencies between the nodes.
However, adding such edges requires extra processing of ASTs and may introduce noise for different learning tasks since there is no consensus on which types of edges are needed for which tasks.

There also exist deep learning techniques that process code syntax trees or abstract syntax trees (ASTs)~\citep{mou2016convolutional,Alon2019,Zhang2019}. 
However, they are limited in how they represent and learn ASTs although ASTs entail all code semantics.
Tree-Based Convolutional Neural Network (TBCNN)~\citep{mou2016convolutional} shares the same computational principle as GGNN, i.e., information is accumulated from nearby children to parent nodes only, which limits the number of iterations for a node to accumulate information from its distant descendants.
Code2vec~\citep{Alon2019} decomposes trees into a bag of path-contexts for learning; ASTNN~\citep{Zhang2019} splits big trees for programs and functions into smaller subtrees for individual statements. They adapt recurrent neural network models to learn the path-contexts or flattened subtrees, but still likely miss code dependency information that is not represented in the decomposed paths and subtrees.


It is desirable to learn code models via ASTs because trees can be more efficiently and precisely constructed from code than graphs without the need of semantic analysis that may be expensive or inaccurate.
Towards this goal, this paper proposes a novel architecture called {\bf TreeCaps} by fusing capsule networks~\citep{sabour2017dynamic} with TBCNN
to build code models from trees, as a complement to graph-based models.
TreeCaps first adapts TBCNN to take in trees and extract (local) node features with its convolution capability and converts the node features into capsules in its {\em Primary Variable Capsule} (PVC) layer where the number of capsules can change for different tree inputs.
It then adapts CapsNet by introducing two methods to route the dynamic number of capsules in PVC to a static number of capsules in its {\em Secondary Capsule} (SC) layer.
Our first method inherits the dynamic routing algorithm~\citep{sabour2017dynamic} for static numbers of capsules; it shares a global transformation matrix across every pair of capsules between the layers~\citep{zhao2018investigating, XinyiC19}.
Our second method is a novel {\em Variable-to-Static} (VTS) routing algorithm that selects the capsules with the most prominent outputs in the PVC layer and squeezes them into a fixed set of capsules. The method utilizes the common intuition that code semantics can often be determined by considering only a portion of code elements.
Further, we apply a dynamic routing algorithm from the capsules in the SC layer to the final {\em Code Capsule} (CC) layer whose number of capsules is fixed according to a specific learning task, to get the vector embeddings of the trees for the task. Compared to the max-pooling method to combine node features in TBCNN, the pipeline of our routing methods (PVC $\rightarrow$ SC $\rightarrow$ CC) can learn more complex combinations of AST features.

Across codebases in C/C++ and Java
with respect to commonly compared program comprehension tasks such as code functionality classification and function name prediction, 
our empirical evaluation shows that TreeCaps achieves better classification accuracy and better F1 score in prediction
compared to other code learning techniques
such as 
Code2vec, 
Code2seq,
ASTNN, 
TBCNN, 
GGNN, GREAT and GNN-FiLM.
We have also applied three types of semantic-preserving transformations~\citep{rabin2020generalizability, zhanggenerating, wang2019learning} that transform programs into syntactically different but semantically equivalent code to attack the models.
Evaluations also show that our TreeCaps models are the most {\em robust}, able to preserve its predictions for transformed programs more than other learning techniques.

\vspace{-6pt}
\section{Related Work}
\label{sec:related}
There has been huge interest in applying deep learning techniques for software engineering tasks such as program functionality classification~\citep{mou2016convolutional, Zhang2019}, function name prediction~\citep{fernandes2018structured}, bug localization~\citep{pradel2018deepbugs, gupta2019neural}, code clone detection~\citep{Zhang2019}, program refactoring~\citep{hu2018deep}, program translation~\citep{chen2018tree}, and code synthesis~\citep{brockschmidt2018generative}. 
A model of source code can often be learned in two steps: (1) convert source code into suitable intermediate representations, and (2) design learning networks to process the representations. 

\citet{mou2016convolutional} parse code into ASTs and design Tree-Based Convolutional Neural Networks (TBCNNs) as the learning networks.
\citet{Allamanis2018} extend ASTs to graphs by adding a variety of code dependencies as edges among tree nodes, intended to represent code semantics, and apply Gated Graph Neural Networks (GGNN) \citep{Li2016} to learn the graphs, which indeed enhances the performance of TBCNN~\citep{mou2016convolutional} for certain tasks. GNN-FiLM~\cite{brockschmidt2019gnn} is also a graph-based model that explores by applying feature-wise linear modulation (FiLM) on Graph Neural Network (GNN). \citet{hellendoorn2019global} proposes a hybrid approach to combine sequence-based models (Recurrent Neural Networks, Transformer) and graph-based models (GNNs) into a model called Graph Relational Embedding Attention Transformer (GREAT) to address the major drawback of GGNN that can only capture local information of the source code.
While the graph-based model extracts local features of source code, the sequence-based model captures global features, their combination improves the performance of GREAT over GNNs.

Code2vec~\citep{Alon2019}, Code2seq~\citep{alon2018code2seq}, and ASTNN~\citep{Zhang2019} are designed based on splitting ASTs into smaller ones, either as a bag of path-contexts or as flattened subtrees representing individual statements, and use various kinds of Recurrent Neural Network (RNN) to learn such code representations. 
Inst2vec \citep{ben2018neural} uses the RNN to model the Intermediate Representation of the binary code that is independent of the source programming language. 

Capsule networks (CapsNet)~\citep{sabour2017dynamic,hinton2018matrix} use dynamic routing to model spatial and hierarchical relations among objects in an image. The techniques have been successfully applied to image processing tasks such as image classification, character recognition, and text classification~\citep{Jayasundara2019,Rajasegaran2019,zhao2018investigating, li2019capsule}. 
However, none of the studies has considered complex tree data as input, which is however natural for programs.
Capsule Graph Neural Networks~\citep{XinyiC19} proposed to classify biological and social network graphs does not handle tree- or graph-based code syntax. 
To the best of our knowledge, we are the first to adapt capsule networks for program source code processing to
learn code models on syntax trees directly, without the need for extra static program semantic analysis techniques that may be expensive or introduce inaccuracies~\citep{Nielson1999}.
\vspace{-6pt}
\section{Tree-based Capsule Networks}
\label{sec:overview}

\begin{figure*}[!h]
	\centering
	\includegraphics[scale=0.28]{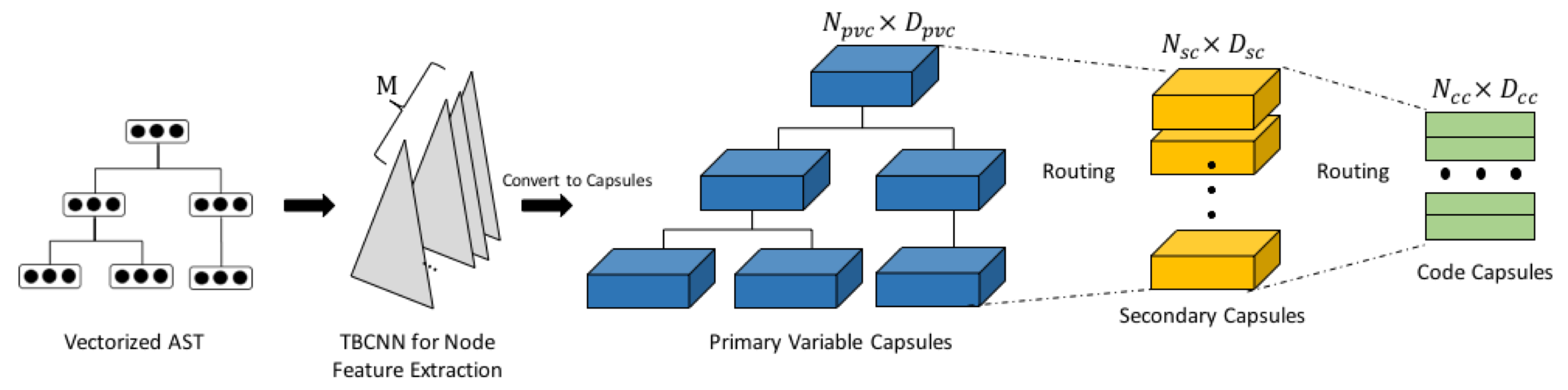}
	\caption{Source codes are parsed, vectorized and fed into the TBCNN to extract node features, then the node features are combined through the TreeCaps network.}
	\label{fig:overview}
\end{figure*}

An overview of the TreeCaps architecture is shown in Fig.~\ref{fig:overview}. 
The steps of our technique are as follows:
\begin{itemize}[leftmargin=*]
	\item The code snippet in the training data is parsed into an AST and vectorized. The node vectors are fed into the TBCNN to extract node features.
	\item The node features will be used as the input for the Primary Variable Capsule (PVC) layer to group the tensor outputs of the TBCNN layers into a set of capsules. The number of capsules in this layer is dynamic
	\item The capsules in the PVC layer are then routed and reduced to a fixed number of capsules in the Secondary Capsule (SC) layer. The SC layer is to combine the capsules in the PVC layer into a new set of capsules, in which the number of capsules in this layer is static.
	\item The outputs of the SC layer are routed to the final Code Capsule (CC) layer where capsules can be seen as the vector representations for the input code, and can be trained with respect to various code comprehension tasks, such as code functionality classification and function name prediction.
\end{itemize}



\vspace{-5pt}
\subsection{Tree-based Convolutional Neural Networks}
We briefly introduce the Tree-based Convolutional Neural Networks (TBCNN, \cite{mou2016convolutional}) for processing tree-structured inputs used in TreeCaps. 


A tree $T = (V, E, X)$ consists of a set of nodes $V$, a set of node features $X$, and a set of edges $E$. An edge in a tree connects a node and its children. 
Each node in an AST also contains its corresponding texts (or tokens) and its type (e.g., operator types, statement types, function types, etc.) from the underlying code.
Initially, we annotate each node $v \in V$ with a $D$-dimensional real-valued vector $x_v \in \mathbb{R}^D$ representing the features of the node. We associate every node $v$ with a hidden state vector $h_v$, initialized from the feature embedding $x_v$,
which can be computed from a simple concatenation of the embeddings of its texts and type~\citep{Allamanis2018}.
The embedding matrices for the texts and types can be learned in the whole model training pipeline. 

In TBCNN, a convolution window over an AST is emulated via a binary tree, where the weight matrix for each node is a weighted sum of three fixed matrices $\mathbf{W}^t$, $\mathbf{W}^l$, $\mathbf{W}^r \in \mathbb{R}^{D \times D}$
(for the ``top'', ``left'', and ``right'' node respectively)
and a bias term $\mathbf{b} \in \mathbb{R}^{D}$.
Hence, for a convolution window of depth $d$ in the original AST containing $K = 2^d -1$ nodes (including the parent node) with vectors $[\mathbf{x}_1, ... , \mathbf{x}_{K}]$, where $\mathbf{x_{i}}\in \mathbb{R}^{D}$,
the convolutional output $\mathbf{y}$ of the window is defined as follows:
\begin{equation}
\mathbf{y} = tanh(\sum_{i=1}^{K}[\eta_{i}^{t}\mathbf{W}^{t}+\eta_{i}^{l}\mathbf{W}^{l}+\eta^{r}_{i}\mathbf{W}^{r}]\mathbf{x}_{i}+\mathbf{b}),
\end{equation}

where $\eta_{i}^{t},\eta_{i}^{l},\eta_{i}^{r}$ are weights calculated corresponding to the depth and the position of the nodes.
One can see this as a way to learn the position of a node inside a tree.
A TBCNN model usually stacks $m$ such convolutional layers together to generate the final node embeddings, where the output at layer $m$ will be used as the input for the next, i.e. the $m + 1$-th layer. 
Each layer has its own $\mathbf{W}^t$, $\mathbf{W}^l$, $\mathbf{W}^r \in \mathbb{R}^{D \times D}$ and the bias term $\mathbf{b} \in \mathbb{R}^{D}$ with different initialization.

\vspace{-5pt}

\subsection{The Primary Variable Capsule Layer (PVC)}

The PVC layer is to group the outputs of the convolutional layers into the set of capsules for the routing purpose. Each convolutional layer will output a tensor with shape $|V| \times D$, where $|V|$ is the number of nodes in the AST, $D$ is the dimension size of the node embedding. There are m such TBCNN layers; then the outputs of such m layers will be a tensor with shape $|V| \times D \times m$. We set $N_{pvc} = |V| \times D$, $D_{pvc} = m$ so that the PVC layer will receive the input of the shape $(N_{pvc} \times D_{pvc})$. It will go through a non-linear squash function \citep{sabour2017dynamic} and get the output with the same shape $(N_{pvc} \times D_{pvc})$. Each output capsule $\mathbf{u}_i$ from the squash function represents the probability of existence of an entity by the vector length, formally defined as:

\begin{equation}
\mathbf{u}_i = \frac{||\mathbf{c}_i||^2}{||\mathbf{c}_i||^2+1} \cdot \frac{\mathbf{c}_i}{||\mathbf{c}_i||}.
\end{equation}

Hence, the output of the PVC layer is $\mathbf{X_{pvc}}\in \mathbb{R}^{N_{pvc}\times D_{pvc}}$.

\subsection{The Secondary Capsule Layer (SC)}
\label{sec:sclayer}

Because $N_{pvc}$ is dynamic as $|V|$ is dynamic, one can not route the output of the PVC layer directly into the final capsule layer (similar to ~\citet{sabour2017dynamic}. To address this, we propose 2 methods to combine the dynamic number of capsules in PVC into static number of capsules in an intermediate layer, called the Secondary Capsule layer. 

\subsubsection{Sharing Weights across Child Capsules with Dynamic Routing (DRSW)}
To combine the capsules in layer $l$ into layer $l+1$, the key is to define a set of transformation matrices. Each matrix is multiplied with each of the capsule in layer $l$ (~\citet{sabour2017dynamic}). In this way, the matrices will be learned as parameters through the end-to-end learning process so the capsules in layer $l$ will be combined through matrices into the capsules in layer $l+1$.
Since the number of capsules in the PVC is dynamic, a global transformation matrix cannot be defined in practice with variable dimensions. The solution for this problem is to defined a shared transformation matrix $\mathbf{W_{s}}\in \mathbb{R}^{N_{pvc}\times D_{pvc} \times D_{sc}}$ across the child capsules, where $N_{pvc}$ is the number of capsules
in the PVC layer \citep{zhao2018investigating}, $D_{sc}$ is the dimension of the capsules in the SC layer, and a dynamic algorithm routes the capsules (as summarized in Algo.\ref{algo:dynamic_routing}).

In Algo.\ref{algo:dynamic_routing}, for each capsule $i$ in the $l$-th PVC layer and each capsule $j$ in the $l+1$-th SC layer, we multiply the output of the PVC layer $\mathbf{u}_i$ by the \textit{shared} transformation matrix $\mathbf{W}_{s}$ to produce the prediction vectors $\mathbf{\hat{u}}_{j|i} = \mathbf{W}_{s}\mathbf{u}_i$. The ``prediction vectors'' are responsible for predicting the strength of each capsule in the PVC layer, then a weighted sum over all
``prediction vectors'' $\mathbf{\hat{u}}_{j|i}$ will produce the capsule $j$ in the SC layer.
The trainable shared transformation matrix learns the part-whole relationships between the primary capsules and secondary capsules, while effectively transforms $\mathbf{u}_i$'s into the same dimensionality as $\mathbf{v}_j$ where each $\mathbf{v}_j$ denotes the capsule output of the SC layer.
The coupling coefficients  $\beta_{ij}$ between capsule $i$ and all the capsules in the SC layer sum to 1 and are determined by a ``routing softmax'' whose initial logits $\alpha_{ij}$ are the log prior probabilities that capsule $i$ in PVC layer should be coupled to capsule $j$ in the SC layer. Then we use $r$ iterations to refine  $\beta_{ij}$ based on the agreements between the prediction vectors $\mathbf{\hat{u}}_{j|i}$ and the secondary capsule outputs $\mathbf{v}_j$ where $\mathbf{v}_j = squash(\sum_{i}^{}\beta_{ij}\mathbf{\hat{u}}_{j|i})$. 

\begin{algorithm}[!h]
	\small
	\caption{Dynamic Routing}\label{algo:dynamic_routing}
	\begin{algorithmic}[1]
		\Procedure{Routing}{}($\mathbf{\hat{u}}_{j|i},r,l$)
		\State Initialize $\forall i \in [1,l], \forall j \in [1,l+1], \alpha_{ij} \leftarrow 0$ 
		\For {r iterations}
		\State $\forall i \in [1,l], \mathbf{\beta}_{i} \leftarrow softmax(\mathbf{\alpha}_{i})$
		\State $\forall j \in [1,l+1], \mathbf{v}_j \leftarrow squash(\sum_{i}^{}\beta_{ij}\mathbf{\hat{u}}_{j|i})$ 
		\State $\forall i \in [1,l], \forall j \in [1,l+1], \alpha_{ij} \leftarrow \alpha_{ij} + \mathbf{\hat{u}}_{j|i} \cdot \mathbf{v}_j$ 
		
		\EndFor
		\State\Return $\mathbf{v}_j$
		
		\EndProcedure
	\end{algorithmic}
\end{algorithm}
\subsubsection{Variable-to-Static Routing (VTS)}


Sharing the transformation matrix reduces the ability to learn different features because each pair of capsules is supposed to have its transformation matrix. Due to this limitation, we offer the second solution to route the variable number of capsules in the PVC layer. 
It is based on an observation of source code that, in practice, not every node of the AST contributes towards a source code learning task. Often, source code consists of non-essential entities, and only a portion of all entities determine the code class. Therefore, we propose a novel variable-to-static capsule routing algorithm, summarized in Algo.~\ref{algo:vts_routing}. The intuition of this algorithm is that we squeeze the variable number of capsules in the PVC layer to a static number of capsules by choosing only the most important capsules in the PVC layer. The major difference between the VTS algorithm and the DRSW algorithm is that the DRSW needs to produce prediction vectors by multiplying the capsule outputs in PVC layer with the shared transformation matrix, and then the prediction vectors will be combined to produce the capsules for SC layer; whereas in the VTS, the capsule outputs in the PVC layer are selected and the prominent ones are used to initialize the capsules in SC layer directly. 

\newcommand*{\skipnumber}[2][1]{{\renewcommand*{\alglinenumber}[1]{}\State #2}\addtocounter{ALG@line}{-#1}}
\begin{algorithm}[!t]
	\small
	\caption{Variable-to-Static Capsule Routing}\label{algo:vts_routing}
	\begin{algorithmic}[1]
		\Procedure{Routing}{}($\mathbf{u}_i,r,a,b$)
		\State $\mathbf{U_{sorted}}\leftarrow sort([\mathbf{u}_1,...,\mathbf{u}_b])$
		\State Initialize $\mathbf{v}_j$ : $\forall i,j \leq a, \mathbf{v_j \leftarrow U_{sorted}}[i]$
		\State Initialize $\alpha_{ij}$ : $\forall j \in [1,a], \forall i \in [1,b], \alpha_{ij} \leftarrow 0$ 
		\For {$r$ iterations}
		\State $\forall j \in [1,a], \forall i \in [1,b], f_{ij} \leftarrow \mathbf{u}_i \cdot \mathbf{v}_j$
		\State $\forall j \in [1,a], \forall i \in [1,b], \alpha_{ij} \leftarrow \alpha_{ij} + f_{ij}$ 
		\State $ \forall i \in [1,b], \bm{\beta}_{i} \leftarrow Softmax(\bm{\alpha}_{i})$
		\State $\forall j \in [1,a], \mathbf{v}_j \leftarrow Squash(\sum_{i}$$ \beta_{ij} \mathbf{u}_i)$ 
		\EndFor
		\State\Return $\mathbf{v}_j$
		
		\EndProcedure
	\end{algorithmic}
\end{algorithm}
We initialize the outputs of the SC layer with the outputs of the $a$ capsules with the highest $L_2$ norms in the PVC layer. Hence, the outputs of the PVC layer, $[\mathbf{u}_1,...,\mathbf{u}_{N_{pvc}}]$, are first ordered by their $L_2$ norms to obtain $\mathbf{U_{sorted}}$, and then the first $a$ vectors of $\mathbf{U_{sorted}}$ are assigned as $\mathbf{v}_j, j \leq a$. 

Since the probability of the existence of an entity is denoted by the length of the capsule output vector ($L_2$ norm), we only consider the entities with the highest existence probabilities for initialization (in other words, highest activation) following the aforementioned intuition.
It should be noted that the capsules with the $a$-highest norms are used {\em only for the initialization}; the actual outputs of the static capsules in the SC layer are determined by iterative runs of the variable-to-static routing algorithm.
It is the capsules with the most prominent outputs along with the capsules of the highest vector similarities to them that get routed to the next layer. In this way, rare capsules, when they have prominent outputs, are still preserved and routed to the next layer.

Next, we route all $b$ capsules
in the PVC layer based on the similarity among them and the static capsule layer outputs.
We initialize the routing coefficients as $\alpha_{ij} = 0$, equally to 
the $b$ capsules in the PVC layer. Subsequently, they are iteratively refined based on the {\em agreement} between the current SC layer outputs $\mathbf{v}_j$ and the PVC layer outputs $\mathbf{u}_i$. The agreement, in this case, is measured by the dot product, $f_{ij} \leftarrow \mathbf{u}_i \cdot \mathbf{v}_j$, and the routing coefficients are adjusted with $f_{ij}$ accordingly.
If a capsule $u$ in the PVC layer has a strong agreement with a capsule $j$ in the SC layer, then $f_{ij}$ will be positively large; whereas if there is strong disagreement, then $f_{ij}$ will be negatively large. Subsequently, the sum of vectors $\mathbf{u}_i$ is weighted by the updated $\beta_{ij}$ to calculate $\mathbf{s}_j$, which is then squashed to update $\mathbf{v}_j$. 

\subsection{The Code Capsules Layer (CC)}

The CC layer outputs the vector embeddings for the code $\mathbf{X_{cc}}\in \mathbb{R}^{N_{cc}\times D_{cc}}$, where $D_{cc}$ is the dimensionality of each code capsule and $N_{cc}$ is fixed with respect to a specific code learning task.
Note in the outputs of the SC layer $\mathbf{X_{sc}}\in \mathbb{R}^{N_{sc}\times D_{sc}}$, $N_{sc}$ is also fixed,

The following subsections explain how we set $N_{cc}$ and train the TreeCaps models for different code learning tasks.

\subsubsection{Code (Functionality) Classification}


This task is to, given a piece of code, classify the functionality class it belongs to.
We want $N_{cc}$ capsules in the CC layer, each of which corresponds to a functionality class of code that appeared in the training data.
As such, we let $N {cc} = \kappa$, where $\kappa$ is the number of functionality classes. 
We calculate the probability of the existence of each class by obtaining $L_2$ norm of each capsule output vector. We use the margin loss \citep{sabour2017dynamic} as the loss function during training.




\subsubsection{Function (Method) Name Prediction}
\label{sec:unsupervised}
This task is to, given a piece of code (without its function header), predict a meaningful name that reflects the functionality of the code.
For this task, following \citet{Alon2019}'s prediction approach,
we let $N_{cc}$ of the CC layer be $1$, and the output of the only capsule represent the vector for the given piece of code. In this case, the output capsules of the CC layer has the shape of $\mathbf{X_{cc}}\in \mathbb{R}^{1\times D_{cc}}$, which is also the code vector that represents for the code snippet $\boldsymbol{C}$, denoted as $\boldsymbol{v_{C}}$.
The vector embeddings of the function are learn-able parameters, formally defined as $functions\_vocab\in \mathbb{R}^{\left|L\right|\times D_{cc}}$, where $L$ is the set of function names found in the training corpus. The embedding of $function_i$ is row $i$ of $functions\_vocab$. The predicted distribution of the model $q\left (l\right)$ is computed as the (softmax-normalized) dot product between the context vector $\boldsymbol{v_{C}}$ and each of the function embeddings:
\begin{equation}
\scriptsize
for\, l_{i}\in L:\, q\left(l_i\right)=\frac{\exp(\boldsymbol{v_{C}}^T\cdot functions\_vocab_{i})}{\sum_{l_{j}\in L}\exp(\boldsymbol{v_{C}}^T\cdot functions\_vocab_{j})}
\end{equation}

where $q\left(l_i\right)$ is the normalized dot product between the vector of $l_i$ and the code vector $\boldsymbol{v_{C}}$, i.e., the probability that a function name $l_i$ should be assigned to the given code snippet $\boldsymbol{C}$.   We choose $l$ that gives the maximum probability for the snippet $\boldsymbol{v_{C}}$. For training the network, we use cross-entropy as the loss function.

\vspace{-5pt}
\section{Empirical Evaluation}
\label{sec:eval}

\paragraph{General Settings.} 
We use fAST, an efficient parser \citep{fast} to parse code into ASTs in a binary format equivalent to SrcML \citep{srcml};\footnote{\url{https://www.srcml.org/}, 400+ node types for multiple programming languages. We chose SrcML because (1) it provides unified AST representations for various languages such as C/C++/Java,
	and (2) it has an extension SrcSlice (\url{https://github.com/srcML/srcSlice})
	to help identify dependencies and construct the graphs needed for GGNN, which is an evaluation baseline.}
we also use another parser PycParser\footnote{\url{https://github.com/eliben/pycparser/}, 50+ node types for C.} used by TBCNN and ASTNN for a fairer comparison and evaluate the effects of parser choices.
For the parameters in our TBCNN layer, we follow \citet{mou2016convolutional} to set the size of type embeddings to 128, the size of text embeddings to 128, and the number of convolutional steps $m$ to 8.
For the capsule layers, we set $N_{sc} = 100$, $D_{sc} = 16$,  $D_{cc} = 16$ and routing iterations $r$ = 3.
We use Tensorflow libraries to implement TreeCaps.
To train the models, we use the Rectified Adam (RAdam) optimizer \citep{liu2019variance} with an initial learning rate of $0.001$ subjected to decay on an Nvidia Tesla P100 GPU. 

\vspace{-5pt}
\paragraph{Baselines} We choose a few
recent code modeling techniques
to compare with TreeCaps: Code2vec~\cite{Alon2019}, Code2seq~\cite{alon2018code2seq}, TBCNN~\cite{mou2016convolutional},
ASTNN~\cite{Zhang2019},
GGNN~\cite{Allamanis2018},
GREAT~\cite{hellendoorn2019global}. We also include a token-based baseline by treating source code as sequences of tokens and using a neural machine translation (NMT) baseline, which is a 2-layer Bi-LSTM, to process the token sequences. A common setting used among all these techniques is that they all utilize both node type and token information to initialize a node in ASTs.
We set both the dimensionality of type embeddings and text embeddings to 128. Note that we try our best to make the baselines as strong as possible by choosing the hyper-parameters above as the ``optimal settings'' according to their papers or code.\footnote{The settings for each of the baselines and parameter analyses can be looked up in the supplementary materials.}

We use different baselines for the two tasks since not all the models were designed for both tasks. 
For the graph-based models (GGNN, GREAT), 
there is no publicly available tool to generate the needed graph representations of code by adding semantic edges into the ASTs as presented in \cite{Allamanis2018}, so we have implemented a tool by ourselves to represent the code as graphs with the assistance of SrcSlice and SrcML.
We include as many edges presented in \cite{Allamanis2018} as possible to ensure the graph-based baselines are strong.
The set of edges we used are: {\texttt{parent\_child, next\_token, last\_lexical\_use, last\_write, return\_to, compute\_from, guarded\_by, guarded\_by\_negation}}. We also add the backward edges for these edge types.
For Code2vec, we follow the settings suggested in their latest Code2seq paper as well as the implementation in the official software artifacts to reproduce their results.
%
%

\subsection{Setups for Code Classification}
\label{lab:set_up_code_classification}

\paragraph{Datasets, Metrics, and Models.}

We use datasets in two different programming languages.
The first Sorting Algorithms (SA) dataset is from~\citet{Bui2019}, which contains 10 algorithm classes of 1000 sorting programs written in {\tt Java}.
The second OJ dataset is from~\citet{mou2016convolutional}, which contains 52000 {\tt C} programs of 104 classes.
We split each dataset into training, testing, and validation sets by the ratios of 70/20/10.
We use the same classification accuracy metric as \citet{mou2016convolutional} for comparing classification results. 

We compare TreeCaps with other techniques that have been applied to the code classification task, such as TBCNN~\cite{mou2016convolutional}, ASTNN~\cite{Zhang2019}, Code2vec~\cite{Alon2019}, GGNNs~\cite{Allamanis2018, fernandes2018structured}.  
Since TBCNN \citep{mou2016convolutional} and ASTNN \citep{Zhang2019} use PycParser to parse code into AST,
we also compare TreeCaps with all the baselines by using both PycParser and SrcML. We also include an \textit{\textbf{ablation study}} to measure the impact of different combinations of node initialization and representation.


\begin{table*}[t]
	\centering
	\caption{Performance in Code Functionality Classification compared. A `-' means that the model is not suited to use the relevant node representation or the parser and thus not evaluated.}
	\label{tab:pc_results}
	\fontsize{7.4}{8.5}\selectfont 
	\begin{tabular}{|c|c|c|c|c|c|c|c|c|c|}
		\hline
		\textbf{Model}        & \multicolumn{3}{c|}{\textbf{SA Dataset (1000 samples)}}           & \multicolumn{6}{c|}{\textbf{OJ Dataset (52000 samples)}}                                                               \\ \hline
		\textbf{Parser}       & \multicolumn{3}{c|}{\textbf{SrcML}}               & \multicolumn{3}{c|}{\textbf{PycParser}}           & \multicolumn{3}{c|}{\textbf{SrcML}}               \\ \hline
		\textbf{Initial Info} & \textbf{Type} & \textbf{Token} & \textbf{Combine} & \textbf{Type} & \textbf{Token} & \textbf{Combine} & \textbf{Type} & \textbf{Token} & \textbf{Combine} \\ \hline\hline\hline
		2-layer Bi-LSTM       & -             & 81.83          & -                & -             & 83.51          & -                & -             & 83.51          & -                \\ \hline
		Code2vec                 & -         & -          & 80.44            & -          & -          & 86.21            & -         & -          & 80.15            \\ \hline
		TBCNN                 & 78.09         & 71.23          & 82.02            & 92.64          & 87.97          & 95.21            & 81.15         & 71.15          & 83.90            \\ \hline
		ASTNN                 & -             & -              & 84.32            & -             & -              & 98.2             & -             & -              & 85.32            \\ \hline
		GGNN                  & 82.12         & 74.25          & 83.81            & -         & -          & -            & 85.23         & 72.23          & 85.89            \\ \hline
		Treecaps-DRSW         & 83.15         & 74.56          & 84.57            & 94.75         & 89.42          & 96.74            & 83.59         & 77.59          & 87.77            \\ \hline
		Treecaps-VTS        & 84.60         & 78.15          & \textbf{85.43}   & 95.88         & 90.21          & \textbf{98.32}   & 83.40         & 79.56          & \textbf{88.40}   \\ \hline
	\end{tabular}
\end{table*}

\vspace{-8pt}
\paragraph{Code Classification Results.}

%

As shown in Table~\ref{tab:pc_results}, TreeCaps models, especially TreeCaps-VTS, have the highest classification accuracy when combining node type and node token information, for both of the SA and OJ datasets.
When only node token information is used, the simpler 2-layer Bi-LSTM models may achieve higher accuracy. 
The OJ dataset also shows that the choice of a parser affects the performance significantly.
The models using PycParser all achieve higher accuracy than the models using SrcML.
This is due to the reason that ASTs generated by PycPaser have only around 50 node types, while SrcML has more than 400 node types, which makes it harder for the networks to learn.
Across the datasets, The TreeCaps-VTS performs consistently the best in terms of the F1 measure among the baselines under different settings. 

\subsection{Setups for Function Name Prediction}

\subsubsection{Datasets, Metrics, and Models.} We use the datasets from Code2seq\cite{alon2018code2seq} containing three sets of Java programs: Java-Small (700k samples), Java-Med (4M samples), and Java-Large(16M samples). These datasets have been split into training/testing/validation by projects.
%
We measure prediction performance using precision (P), recall (R), and F1 scores over the sub-words in generated names, following the metrics used by \citet{Alon2019, fernandes2018structured}.
For example, a predicted name \texttt{result\_compute} is considered to be an exact match of the actual name \texttt{computeResult}; predicted \texttt{compute} has full precision but only 50\% recall; and predicted \texttt{compute\_model\_result} has full recall but only 67\% precision.

We use these baselines for the function name prediction task:
Code2vec,
TBCNN,
Code2seq,
GGNN,
the 2-layer Bi-LSTM, and  
and GREAT~\cite{hellendoorn2019global}, a hybrid model mixing sequence-based and graph-based techniques.
The inputs for GREAT is graph representations of code, similar to GGNN, and we have adapted this baseline into the function name prediction task.
We train each of the models for 50 epochs for each of the three datasets. We also measure the training time for each of the models.

\subsubsection{Function Name Prediction Results.} As shown in Table \ref{tab:method_name_results}, TreeCaps-VTS outperforms all other baselines for most of the settings.
TreeCaps-DRSW also performs well but still worse than TreeCaps-VTS. 
Both TreeCaps-VTS and TreeCaps-DRSW 
are better than the graph-based models (GGNN, GREAT) and path-based model (Code2seq, Code2vec\footnote{Noted that the results for Code2vec reported \citet{Alon2019} is different from the Code2vec results reported in our paper. This inconsistency has appeared in Code2seq, which is the later work of the same group of the author of Code2vec and has been explained in the rebuttal phase of Code2seq at ICLR'19 (see \url{https://openreview.net/forum?id=H1gKYo09tX}). }) without the need for additional code dependency analysis for constructing graphs.
Regarding the training time, GGNN is the longest. The training time of both TreeCaps models is comparable to GREAT, the state-of-the-art graph-based technique to model source code, while TreeCaps-VTS is slightly faster.

\begin{table*}[t]
	\centering
	\caption{Performance of TreeCaps and the baselines for Function (Method) Name Prediction}
	\label{tab:method_name_results}
	\fontsize{7.5}{8.5}\selectfont 
	\begin{tabular}{|c|c|c|c|c|c|c|c|c|c|c|c|c|}
		\hline
		\textbf{Model}  & \multicolumn{4}{c|}{\textbf{java-small (700k Samples)}}                   & \multicolumn{4}{c|}{\textbf{java-med (4M Samples)}}                       & \multicolumn{4}{c|}{\textbf{java-large (16M Samples)}}                 \\ \hline
		\textbf{Metric} & \textbf{P}     & \textbf{R}     & \textbf{F1}    & \textbf{Training Time} & \textbf{P}     & \textbf{R}     & \textbf{F1}    & \textbf{Training Time} & \textbf{P}     & \textbf{R}     & \textbf{F1} & \textbf{Training Time} \\ \hline
		2-layer Bi-LSTM  & 40.02          & 31.84          & 35.46          & 26.3h                  & 49.73          & 40.12          & 44.82          & 65.2h                  & 56.56          & 49.27          & 52.63       & 150h                   \\ \hline
		TBCNN           & 40.89          & 27.67          & 32.24          & 20.6h                  & 45.23          & 41.41          & 43.23          & 58.7h                  & 58.15          & 40.91          & 49.40       & 165h                   \\ \hline
		Code2vec        & 23.35          & 22.01          & 21.36          & 47.9h                  & 36.43          & 27.93          & 31.89          & 91.6h                  & 44.24          & 38.25          & 41.56       & 222h                   \\ \hline
		
		Code2seq        & 50.42          & 35.43          & 42.56          & 56.3h                  & 62.56          & 46.83          & 53.66          & 100h                  & 63.25          & 54.03          & 58.96       & 235h                   \\ \hline
		
		GGNN            & 40.25          & 35.25          & 36.86          & 75.8h                  & 50.14          & 41.25          & 45.31          & 142h                   & 50.18 & 44.25          & 46.23       & 280h                   \\ \hline
		GREAT           & 47.25          & 39.97          & 43.56          & 55.5h                  & 57.15          & 44.12          & 51.42          & 110h                   & 61.35          & 55.86          & 58.25       & 205h                   \\ \hline \hline\hline
		TreeCaps-DRSW  & 45.19          & 39.49          & 42.89          & 61.5h                  & 60.19          & 41.15          & 52.56          & 125h                   & 59.41          & 52.93          & 57.82       & 153h                   \\ \hline
		TreeCaps-VTS    & \textbf{52.62} & \textbf{41.36} & \textbf{46.78} & 45.1h                  & \textbf{64.38} & \textbf{48.87} & \textbf{55.67} & 105h                   & \textbf{66.85} & \textbf{56.32} & \textbf{61.34}       & 180h                   \\ \hline
	\end{tabular}
\end{table*}
%

\subsection{Model Analysis}
\begin{table}[t]
	\centering
	\caption{Model robustness, measured as percentage of predictions changed wrt.~semantic-preserving program transformations. The lower the more robust.}
	\label{tab:stability}
	\fontsize{7.4}{8.5}\selectfont 
	\begin{tabular}{|c|c|c|c|l|}
		\hline
		\textbf{Model} & \textbf{VR}        & \textbf{US}               & \textbf{PS}              & \textbf{Average}          \\ \hline\hline\hline
		Code2vec      & 22.45\%         & 19.42\%          & 26.56\%         & 22.81\%          \\ \hline
		Code2seq      & 16.84\%         & 21.82\%          & 20.12\%         & 19.59\%          \\ \hline
		TBCNN         & 11.16\%         & 19.36\%          & 21.67\%         & 17.39\%          \\ \hline
		GGNN          & 15.34\%          & 18.89\% & 16.42\%         & 16.88\%          \\ \hline	
		GREAT          & 13.48\%          & 17.75\% & 16.51\%         & 15.90\%          \\ \hline		
		TreeCaps-DRSW & 11.23\%          & 15.76\%          & 16.54\%         & 14.51\%          \\ \hline
		TreeCaps-VTS  & \textbf{9.53\%} & \textbf{14.08}\%          & \textbf{13.87\%} & \textbf{12.49\%} \\ \hline
	\end{tabular}
\end{table}
\begin{figure*}[t]\centering
	
	\begin{tabular}{@{}cccc@{}}
		\includegraphics[width=1.0\columnwidth]{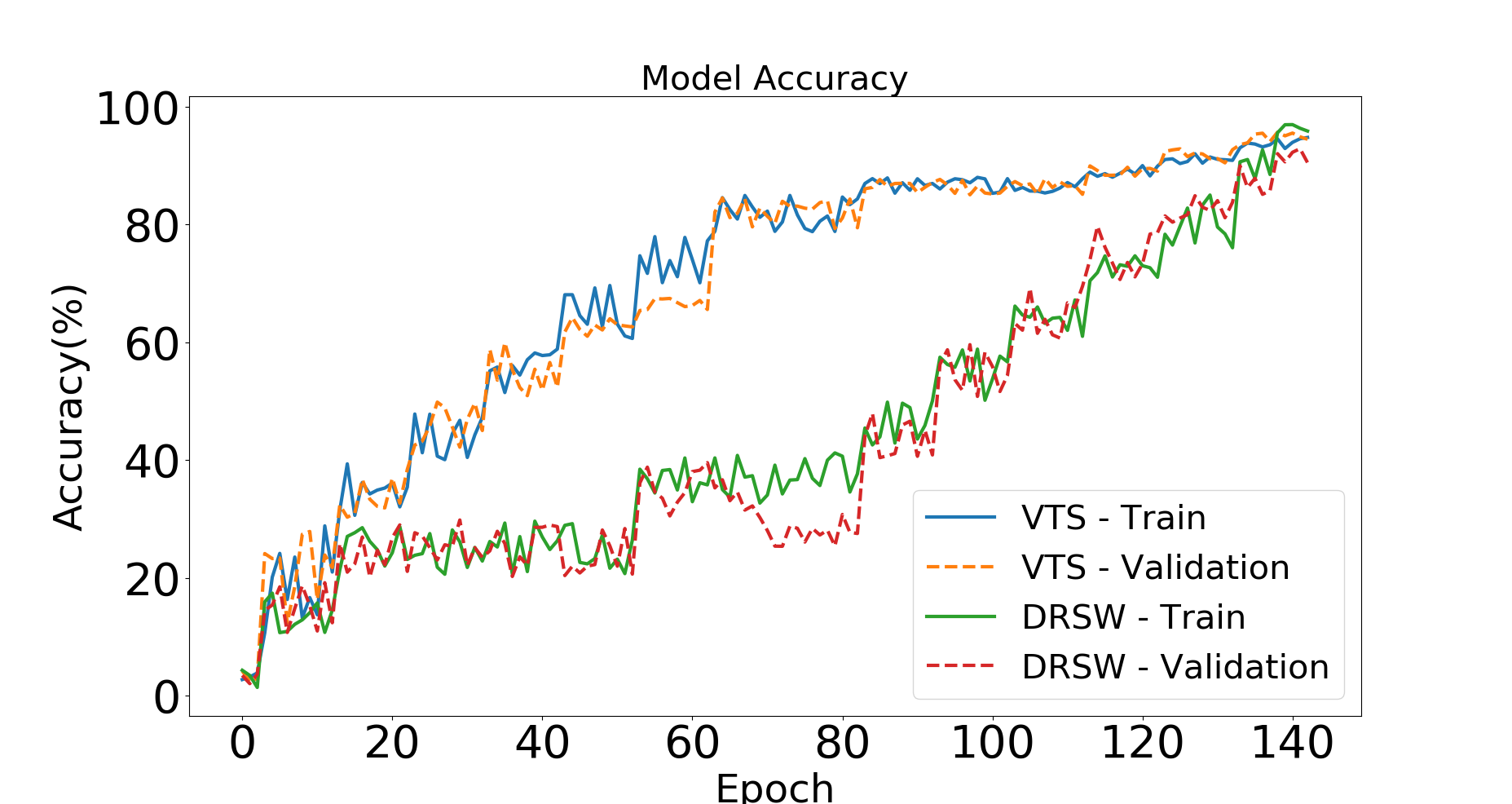} &
		\includegraphics[width=1.0\columnwidth]{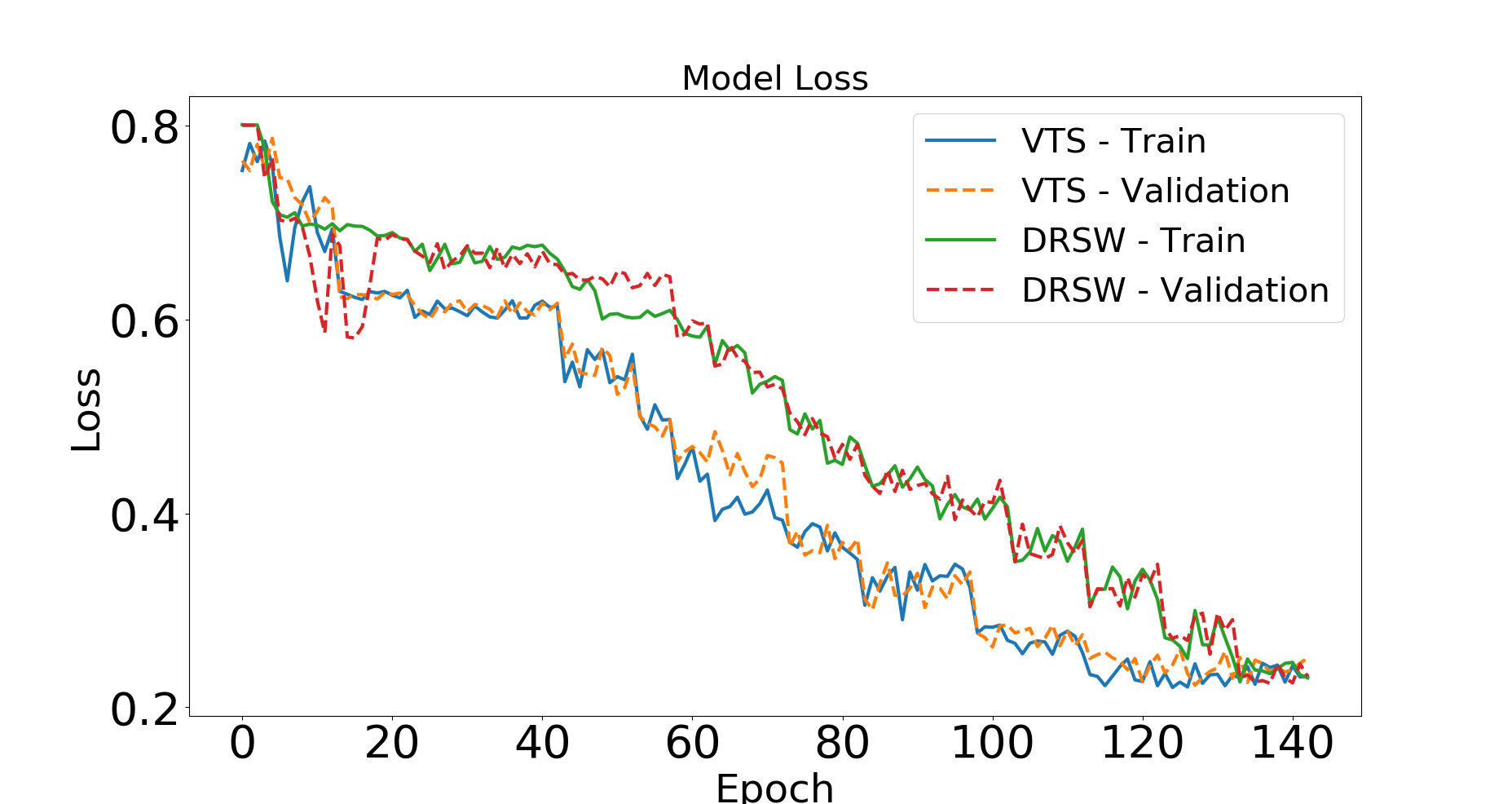}
	\end{tabular}
	\caption{Comparisons between the Two Routing Algorithms}
	\label{tab:comparisons}
\end{figure*}

To better understand the importance of different components of our approach, we evaluate the effect of various aspects of the TreeCaps models. This subsection provides a robustness analysis and a comparison between DRSW algorithm and VTS algorithm\footnote{More analyses on the effect of the SC layer, effect of dimension size of capsules can be found in the supplementary materials.}.

\subsubsection{Robustness of Models}
We measure the robustness of each model by applying the semantically-preserving program transformations to the Java-large’s test set for the \textit{function name prediction} task. We follow \citet{wang2019learning,rabin2020generalizability} to transform programs in three ways that change code syntax but preserve code functionality: (1) Variable Renaming (VN), a refactoring transformation that renames a variable in code, where the new name of the variable is taken randomly from a set of variable vocabulary in the training set;
(2) Unused Statement (US), inserting an unused string declaration to a randomly selected basic block in the code; and
(3) Permute Statement (PS), swapping two independent statements (i.e., with no dependence) in a basic block in the code. 

The Java-large test set is thus transformed into a new test set.
We then examine if the models make the same predictions for the programs after transformation as the prior predictions for the original programs.
We use\textit{ \textbf{percentage of predictions changed}} ($PPC$) as the metric used by \citep{rabin2020generalizability, zhanggenerating, wang2019learning} to measure the robustness of the code models.
Formally, suppose $P$ denotes a set of test programs, a semantic-preserving program transformation $T$ that transforms $P$ into a set of transformed programs $P'=\{p'=T(p)|p\in P\}$, and a source code model $M$ that can make predictions for any program $p$: 
$M(p)=l$, where $l\in L$ denotes a predicted label for $p$ according to a set of labels $L$ learned by $M$,
we compute the percentage of predictions changed as:
\begin{equation}
PPC=\frac{|\{p' \in P'| M(p) \neq M(p')\}}{|\{p' \in P'\}|}*100
\end{equation}
Lower $PPC$ values for $M$ suggest higher robustness as they can maintain more of correct predictions with respect to the transformation. As shown in Table \ref{tab:stability}, 
TreeCaps-VTS is the most robust model against the program transformations.
Although more kinds of program transformations could be applied to evaluate model robustness in our future work, the current analysis gives us the confidence that TreeCaps can be more robust against attacks via adversarial examples \citep{ramakrishnan2020semantic, bielik2020adversarial}.

\subsubsection{Comparison between the Two Routing Algorithms}



%

Figure \ref{tab:comparisons} shows the comparisons between the Dynamic Routing algorithm with Shared Weights (DRSW) and Variable-to-Static Routing algorithm (VTS)
for the \textit{code classification} task on the OJ Dataset.
There are two main observations: (1) when DRSW is used, the loss decreases slower than when VTS is used (in the right plot);
and (2) VTS improves validation accuracy faster than DRSW (in the left chart).
A reason is that DRSW has to learn an additional shared transformation matrix $\mathbf{W_{s}}$, resulting in slower convergence due to a larger number of parameters to be learned.

%

%

\vspace{-5pt}
\section{Discussion} \label{sec:discussion}

\paragraph{Choice of Node Feature Extractor}
For the step to extract the node features, we chose TBCNN because it was designed to process ASTs that usually contain deeper and larger numbers of nodes per code snippet than natural language parse trees per sentence, and it has been shown to outperform TreeLSTM in software engineering tasks such as code classification~\cite{mou2016convolutional} and NLP tasks such as natural language inference~\cite{mou2015discriminative}.

\paragraph{Relationship with Global Model of Source Code}
GREAT~\cite{hellendoorn2019global} is a hybrid approach to combine sequence-based and graph-based model to better capture both local and global features of code. TreeCaps shares the same synergy but with a different approach. The feature extraction step is to extract local features of the source code and the routing mechanism of the capsules is to combine the global features of the source code. We have shown in our evaluation that our capsule-based method performs better than GREAT, both in terms of F1 score and training time. Further research is needed to explore how different types of features are captured inside the capsules.
\\
\vspace{-15pt}
\section{Conclusion} \label{sec:conclusion}
We propose TreeCaps, a novel neural network architecture that incorporates tree-based convolutional neural networks (TBCNN) into capsule networks for better learning of code on abstract syntax trees.
To handle dynamic numbers of capsules produced from TBCNN,
we propose two methods to route the capsules in the Primary Variable Capsule layer to a fixed number of capsules in the Secondary Capsule layer. 
We are the first to re-purpose capsule networks over syntax trees to learn code without the need for explicit semantics analysis.
Our empirical evaluations have shown that TreeCaps can outperform existing code learning models (e.g., Code2vec, TBCNN, ASTNN, GGNN, GREAT, GNN-FiLM) for two different program comprehension tasks (e.g., code functionality classification and function name prediction) on C/C++/Java programs. It is our belief that the new method can be applied to other software engineering tasks such as bug localization and clone detection.
%

A limitation of TreeCaps is similar to the original capsule networks and many other neural networks: it still lacks explainability. Software developers may require additional evidence before accepting the predication results, which suggests future work that relating TreeCaps outputs to certain visible patterns in code could help explain the predictions.

\section{Acknowledgments} \label{sec:ack}

This research is supported by the Singapore Ministry of Education (MOE) Academic Research Fund (AcRF) Tier 1 grant and RISE Lab Operational Fund from SIS at SMU, Singapore MOE AcRF Tier 2 Award No.~MOE2019-T2-1-193, Royal Society International Collaboration projects (Big Code Forensic Analytics in Secure SE IES/R1/191138, IES/R3/193175), EU H2020 EngageKTN project on Safer Drone Flights (https://droneidentity.eu), EPSRC STRIDE (Socio-technical resilience in software development) project (EP/T017465/1), Huawei Trustworthy Lab, Ireland Research Centre. We also thank the anonymous reviewers for their insightful comments and suggestions, and thank the authors of related work for sharing data.

%

	\balance

	\bibliography{references}

\end{document}